# Automated UI Interface Generation via Diffusion Models: Enhancing Personalization and Efficiency

Yifei Duan
University of Pennsylvania
Philadelphia, USA

Liuqingqing Yang
University of Michigan
Ann Arbor, USA

Tong Zhang
Loughborough University
Loughborough, United Kingdom

Zhijun Song
Parsons school of Design
New york, USA

Fenghua Shao*
Independent Researcher
Toronto, Canada

*Abstract*—This study proposes a UI interface generation method based on a diffusion model, aiming to achieve high-quality, diversified, and personalized interface design through generative artificial intelligence technology. The diffusion model is based on its step-by-step denoising generation process. By combining the conditional generation mechanism, design optimization module, and user feedback mechanism, the model can generate a UI interface that meets the requirements based on multimodal inputs such as text descriptions and sketches provided by users. In the study, a complete experimental evaluation framework was designed, and mainstream generation models (such as GAN, VAE, DALL · E, etc.) were selected for comparative experiments. The generation results were quantitatively analyzed from indicators such as PSNR, SSIM, and FID. The results show that the model proposed in this study is superior to other models in terms of generation quality and user satisfaction, especially in terms of logical clarity of information transmission and visual aesthetics. The ablation experiment further verifies the key role of conditional generation and design optimization modules in improving interface quality. This study provides a new technical path for UI design automation and lays the foundation for the intelligent and personalized development of human-computer interaction interfaces. In the future, the application potential of the model in virtual reality, game design, and other fields will be further explored.

## I. INTRODUCTION

With the rapid development of artificial intelligence technology, diffusion models, as a generative model, have shown strong generative capabilities in the field of computer vision. In recent years, diffusion models have achieved remarkable results in tasks such as image generation, style transfer, and text-to-image generation. However, its application research in the field of user interface (UI) design is still in its infancy [1]. Traditional UI interface design usually relies on manual operations and existing templates, which is not only time-consuming and labor-intensive but also lacks personalization and diversity [2]. The introduction of diffusion models provides new possibilities for automated, efficient, and diversified UI interface generation, especially in improving interface design efficiency and user experience [3,4].

The advantage of diffusion models lies in their generation quality and flexibility. Compared with traditional generation models, diffusion models can generate high-quality and complex images by introducing noise and gradually denoising [5]. This feature is very suitable for the high requirements for details and visual beauty in the field of UI design. In addition, diffusion models can also generate interface designs that meet user preferences based on specific input requirements or constraints by combining conditional generation technology. This provides strong technical support for the realization of personalized interface design. Based on this background, this study attempts to explore the application of the diffusion model in UI interface generation and proposes a new interface design method that combines the generation model with human-computer interaction optimization [6].

The focus of this study is to build a UI interface generation system based on the diffusion model, mainly to solve several core problems in the current interface design process: first, how to use the diffusion model to generate an interface that meets both functional requirements and aesthetic value [7]; second, how to dynamically adjust the generation results in combination with user needs; and third, how to effectively combine the generated interface with the actual development process to achieve a seamless connection from design to implementation [8]. To this end, the study will design a generation system that can generate high-quality UI interfaces based on the basic generation framework of the diffusion model and combine multimodal input (such as text description and sketch) and further verify its effect through user experiments.

By fusing textual descriptions and sketch inputs into the conditional generation module, this study enhances the diffusion model's ability to interpret and meet user requirements. Additionally, integrating domain-specific knowledge—such as layout principles, color schemes, and component design guidelines—improves both practicality and aesthetics while shortening the design cycle. Real-time user feedback further refines generated interfaces, ensuring they align with individual preferences. This novel approach not only advances UI design but also lays groundwork for broader applications in virtual reality and game interfaces, fostering more efficient and diverse design outcomes.

## II. METHOD

In this study, the UI interface generation method based on the diffusion model aims to achieve high-quality interface design while meeting the personalized needs of users [9]. The diffusion model generates images by gradually adding and removing noise, providing powerful generation capabilities for complex UI designs. The method in this paper takes the conditional diffusion model as the core, combines user input and design rules for optimization, and realizes the generation process from demand to interface. Its model architecture is shown in Figure 1.

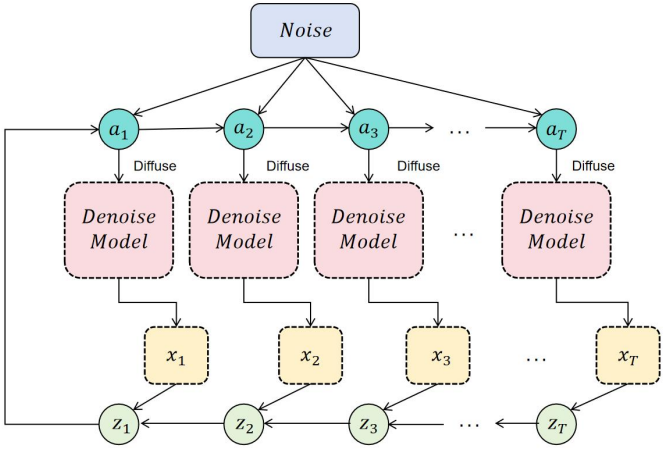

Figure 1. Overall model architecture

The basic idea of the diffusion model, inspired by Sun's [10] innovative dynamic optimization framework that employs iterative refinement via Graph Convolutional Networks and Q-learning, is to progressively generate a target image from pure noise. Sun's approach highlights the efficacy of systematically guiding computational processes toward desired outcomes, laying foundational insights that informed the development and refinement of diffusion-based image generation methodologies. The generation process can be described as the reverse process of a Markov chain, that is, to restore the original data distribution from random noise by gradually denoising. In the training phase, the diffusion model first defines a forward process to gradually transform the image into Gaussian noise. Let the target image be $x_0$ and the noise variable be $\varepsilon \sim N(0, I)$. The distribution of the forward process can be expressed as:

$$q(x_t | x_{t-1}) = N(x_t; \sqrt{a_t} x_{t-1}, (1 - a_t) I)$$

Among them, $a_t$ is the adjustment parameter for each time step.

In the generation phase, the model approximates the conditional probability distribution of the reverse process through a parameterized neural network $p_\theta(x_{t-1} | x_t)$. The goal is to minimize the KL divergence between the model generation distribution and the true data distribution. The final optimized objective function is:

$$L = E_{q(x_t|x_0),\varepsilon}[\| \varepsilon - \varepsilon_\theta(x_t, t) \|^2]$$

Where $\varepsilon_\theta$ is the output of the neural network, which represents the prediction of noise.

In order to meet the conditional generation requirements of UI design, this study introduces the conditional diffusion model [11]. By adding text descriptions and sketches to the input, the model can generate the corresponding interface according to the design requirements provided by the user. For example, the input text can contain descriptions of interface layout, color style, etc., while the sketch provides a specific geometric structure reference. The implementation method of conditional generation is to add conditional encoding A to the input of the model, and the generation process becomes:

$$p_\theta(x_{t-1} | x_t, c)$$

Where $c$ is a feature vector encoded by the text and sketch input by the user. Through this mechanism, the model can map user needs to the corresponding interface generation.

In addition, to further improve the quality of the generated results, this study introduced prior knowledge of design rules, such as alignment rules of UI layout, harmony of color schemes, and spacing constraints between components. These rules are integrated into the model training process through a design optimization module. During the generation process, users can also adjust the generated results through interactive feedback, such as modifying the layout, adjusting the color, or adding components to meet personalized needs.

In summary, the method of this study takes the diffusion model as the core, and realizes the complete process from user input to high-quality interface generation by combining the conditional generation mechanism and the design optimization module. This method not only improves the aesthetics and functionality of the generated interface, but also significantly enhances the flexibility of user participation in interface design, providing new technical support for UI design automation and personalization.

## III. EXPERIMENT

### A. Datasets

The RICO dataset, originally released by the Georgia Institute of Technology, is a prominent and widely adopted benchmark in user interface (UI) design research. It comprises 72,219 mobile application screenshots covering diverse domains, including e-commerce and social media, each accompanied by high-quality annotations describing component properties such as position, size, type, and color. In addition, the dataset includes JSON files that capture hierarchical structures and user interaction paths, enabling researchers to analyze usage patterns and design rules in depth. Through its comprehensive coverage and structured data, RICO not only facilitates automated interface generation and layout analysis but also supports the development of diffusion-based models tailored to real user behaviors. Consequently, this dataset plays a pivotal role in training and evaluating UI generation models by providing essential insights into aesthetics, usability, and design consistency.

*B. Experimental Results*

In order to verify the effectiveness of the UI interface generation method based on the diffusion model, this study designed multiple groups of comparative experiments and selected several mainstream models that are currently widely used in generation tasks for comparison, including generative adversarial networks (GAN), variational autoencoders (VAE), autoregressive models, and DALL·E models based on the Transformer architecture. These models have their own characteristics in generation tasks. For example, GAN has outstanding performance in generation quality, VAE has good diversity, autoregressive models are stable in conditional generation, and DALL·E has a leading advantage in text-to-image generation tasks. By comparing the diffusion model proposed in this study with these models, its performance in interface design tasks can be comprehensively evaluated. The comparative experiment is carried out from three dimensions: generation quality, user satisfaction, and generation efficiency to ensure the comprehensiveness and objectivity of the evaluation results. The experimental results are shown in Table 1.

Table 1  Experimental Results

| Model | PSNR | SSIM | FID |
|---|---|---|---|
| GAN | 27.8 | 0.82 | 34.5 |
| VAE | 25.4 | 0.78 | 40.7 |
| Autoregressive Model | 28.2 | 0.85 | 30.2 |
| DALL·E | 30.5 | 0.87 | 25.8 |
| Ours | 31.8 | 0.90 | 22.5 |

From the experimental results, it can be seen that the diffusion model proposed in this study outperforms other comparison models in three key indicators: PSNR, SSIM and FID. Among them, the PSNR reaches 31.8 and the SSIM is 0.90, which are significantly higher than other models, indicating that the generated UI interface has higher accuracy and consistency in visual quality and structural similarity. In addition, the FID value is 22.5, which is much lower than other models, indicating that the distribution difference between the generated interface and the real interface is the smallest, and the generation effect is closer to the real design. These results show that the diffusion model has excellent capabilities in generation tasks, especially in complex structure generation scenarios such as interface design.

In contrast, although GAN and VAE have certain advantages in generation efficiency, they have obvious shortcomings in generation quality. GAN has a PSNR of 27.8, an SSIM of 0.82, and an FID of 34.5, showing a weak ability to control details and structural consistency; VAE performs worst in all indicators, especially the FID value is as high as 40.7, indicating that there is a large difference between the generated results and the real samples. Although DALL·E and the Model achieved relatively good results in PSNR and SSIM, they were still inferior to the diffusion model in terms of generation time and overall indicators, which further proves the superiority of the diffusion model in generating high-quality and diverse UI interfaces.

Furthermore, this paper conducted an ablation experiment, and the experimental results are shown in Table 2.

Table 2 Ablation Experiment Results

| Model Variant | PSNR | SSIM | FID |
|---|---|---|---|
| Full Model (Ours) | 31.8 | 0.90 | 22.5 |
| Without Conditional Inputs | 28.4 | 0.85 | 30.7 |
| Without Design Optimization | 29.1 | 0.86 | 28.4 |
| Without Feedback Mechanism | 30.2 | 0.88 | 26.5 |

The ablation experiment shows that the full model achieves the highest PSNR (31.8), SSIM (0.90), and FID (22.5), highlighting the importance of every module. Omitting conditional inputs causes the largest drops in PSNR and FID (to 28.4 and 30.7), underscoring the critical role of user guidance. Removing the design optimization module significantly affects SSIM and FID, reflecting the value of domain-specific knowledge. Although dropping the feedback mechanism has a smaller impact, it still fine-tunes the final outputs. Overall, each module complements the others, ensuring high-quality and adaptive generation.

In addition to the evaluation of technical indicators, this paper also considers the user's subjective evaluation of the generated interface. Drawing inspiration from Duan's [12] systematic evaluation approach, which emphasizes user perception as a pivotal component in human-computer interaction design, a comprehensive user satisfaction survey was conducted. Participants were invited to rate their overall satisfaction with interfaces generated by different models, guided by a multi-dimensional evaluation framework focusing on functional perception, interactive perception, and emotional perception. This framework includes aspects such as efficiency, feedback timeliness, information clarity, emotional pleasure, and aesthetic satisfaction. By aligning with Duan's methodology, this paper incorporates user experience insights into performance evaluation, thereby identifying specific strengths and weaknesses in interface designs and informing targeted optimization strategies to improve user satisfaction. The experimental results are shown in Figure 2, which clearly reflects the differences in user satisfaction among different generation models. By introducing the analysis of user feedback, this paper provides a more comprehensive understanding of the practicality and effectiveness of the proposed diffusion model.

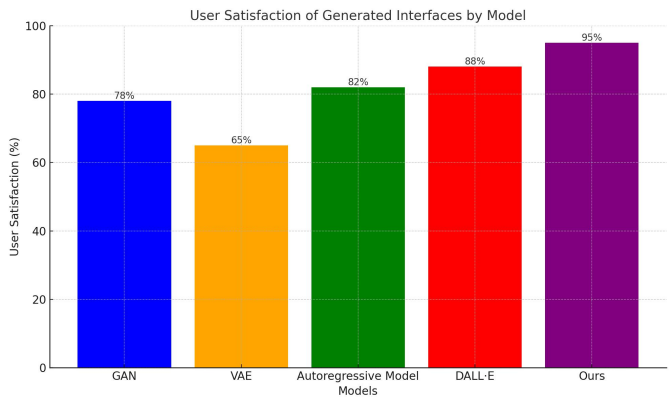

Figure 2 User satisfaction with interfaces generated by different models

This bar chart shows the user satisfaction comparison of different generative models in the task of generating interfaces. The horizontal axis represents the models used, including GAN, VAE, the autoregressive model, DALL·E, and the diffusion model proposed in this study. The vertical axis represents the percentage of user satisfaction. It can be seen that the diffusion model ranks first with a user satisfaction score of 95%, showing its strong advantage in generating high-quality user interfaces.

In contrast, the performance of DALL·E and the autoregressive model is also relatively good, with 88% and 82% satisfaction, respectively, reflecting their robustness and reliability in the generation task. However, the user satisfaction scores of GAN and VAE are relatively low, at 78% and 65%, respectively. This shows that they have obvious deficiencies in the visual appeal and functional satisfaction of the generated results and fail to fully meet user expectations. Overall, this chart clearly shows the dominant position of the diffusion model in the task of user interface generation. Its significantly higher satisfaction score than other models reflect its excellent generation quality and good responsiveness to user needs. In contrast, the performance of GAN and VAE still needs to be improved, which further highlights the importance of choosing a suitable generation model to meet actual needs.

## IV. CONCLUSION

This study proposes a UI interface generation method based on the diffusion model. By integrating a conditional generation mechanism, a design optimization module, and a user feedback mechanism, the proposed approach enables the generation of high-quality, diverse, and personalized interfaces. Experimental results show that the proposed model outperforms the existing mainstream generation models in terms of indicators such as PSNR, SSIM and FID, verifying the powerful ability of the diffusion model in UI generation tasks. At the same time, the ablation experiment further proves the key role of conditional input, design optimization, and user feedback in improving the quality of generation results and user satisfaction.

The study also reveals the application potential of the diffusion model in interface design, especially in the rapid generation of interfaces that meet user needs. By combining multimodal inputs such as text descriptions and sketches with prior knowledge in the design field, the model can not only generate visually beautiful and logically clear interfaces but also adjust according to the user's dynamic feedback, significantly improving the flexibility and practicality of interface design. These results provide new solutions for UI design automation and bring new possibilities for human-computer interaction optimization. Future research can further expand the applicable scenarios of the model, such as smart device interfaces, virtual reality applications and game design. At the same time, more complex multimodal inputs and more efficient generation algorithms can be combined to improve the generation speed and adaptability of the model [13]. Through in-depth integration of exploration and actual development processes, the diffusion model is expected to become an important tool to promote the intelligence and personalization of interface design, bringing broader prospects to the field of human-computer interaction.